# Sentiment Analysis of Social Media Data for Predicting Consumer Behavior Trends Using Machine Learning

S M Rakib Ul Karim[1,] Rownak Ara Rasul[2] & Tunazzina Sultana[3]

[1,2]PhD Fellow, Department of Electronic and Electrical Engineering, University of Missouri, Columbia, USA &
[3]Department of Marketing, University of Chittagong, Bangladesh

## Abstract

In the era of rapid technological advancement, social media platforms such as Twitter (X) have emerged as indispensable tools for gathering consumer insights, capturing diverse opinions, and understanding public attitudes. This research investigates the application of advanced machine learning methods for sentiment analysis on Twitter data, with a particular focus on predicting consumer trends. Using the "Sentiment140" dataset, this study aims to detect evolving patterns and shifts in consumer preferences across product categories, with "car" used as an illustrative example.

The study employs a comprehensive data collection and preparation workflow to cleanse and structure Twitter data for analysis. Advanced machine learning models, including Support Vector Machines (SVM), Naive Bayes, Long Short-Term Memory (LSTM) networks, and Bidirectional Encoder Representations from Transformers (BERT), were utilized to classify sentiments and predict trends. Model performance was evaluated using metrics such as accuracy, precision, recall, and F1 score, with BERT achieving the highest performance (Accuracy: 83.48%, Precision: 79.37%, Recall: 90.60%, F1 Score: 84.61%).

The findings highlight the superior efficacy of models like LSTM and BERT in capturing complex linguistic patterns and contextual relationships within text data. These models not only enhance prediction accuracy but also provide deeper insights into consumer behavior. Temporal analysis revealed sentiment trends across hours, days, and months, while semantic similarity and Named Entity Recognition (NER) identified related terms, brands, and themes, offering actionable insights for businesses.

This study contributes significantly to sentiment analysis research by addressing challenges such as sarcasm detection, multilingual data processing, and the integration of real-time analysis. By providing a scalable framework adaptable to various industries, this research facilitates the generation of precise and practical consumer insights, enabling businesses to navigate the dynamic landscape of social media-driven markets.

# Introduction

In the modern era of digital transformation, social media has evolved into an indispensable medium for communication, information sharing, and expressing viewpoints. Platforms such as Twitter, Facebook, and Instagram have redefined interactions between individuals and businesses, functioning as both personal communication channels and invaluable repositories of data for understanding public attitudes and behaviors. These platforms provide unparalleled opportunities for analyzing consumer sentiment, enabling organizations to derive actionable insights from unstructured text (Gagandeep Kaur, 2023; Stasberger GD, 2023).

The field of natural language processing (NLP) has advanced significantly to address the challenges of extracting subjective information from text, commonly referred to as sentiment analysis. This technique evaluates whether a given text conveys positive, negative, or neutral sentiment. Sentiment analysis has become a cornerstone for businesses aiming to decode consumer preferences and emotions, transforming chaotic datasets into strategic insights (Sibu Skaria, 2024; Geeksforgeeks, 2024)

Unlike traditional market research methods, such as surveys and focus groups, social media data offers immediate and dynamic insights into consumer behavior. With approximately 500 million tweets generated daily by Twitter's 330 million active users, organizations now have access to vast volumes of real-time data (Statista, 2023; Omnicoreagency, 2024). This scale enables the detection of even subtle shifts in public mood, which is crucial for accurately forecasting emerging trends. Social media's interactive nature also allows businesses to promptly address changes in consumer perception, providing a continuous feedback loop that is essential for maintaining a competitive edge in today's fast-paced markets.

Machine learning approaches have revolutionized sentiment analysis by leveraging cutting-edge algorithms and architecture. Recent advancements, particularly between 2020 and 2024, have highlighted the efficacy of deep learning models such as Bidirectional Encoder Representations from Transformers (BERT). These models, pre-trained on extensive text corpora and fine-tuned for specific datasets, excel at capturing nuanced linguistic patterns and contextual relationships, achieving state-of-the-art results across various NLP tasks, including sentiment analysis.

Sentiment analysis is one of the numerous fields of computer research that focus on opinion-oriented natural language processing. Since it deals with the opinion of the users, it also known as opinion mining. Among other things, these opinion-based investigations cover genre differences, mood and emotion recognition, ranking, relevance calculations, text viewpoints, text source identification, and opinion-based summarization. Sentiment analysis has emerged as a fascinating new social media trend with a wide range of useful applications, including subcomponent technology applications (recommender systems), business applications (marketing intelligence), and product and service benchmarking and improvement (Pang, 2008).

Notably, the COVID-19 pandemic accentuated the critical role of sentiment analysis in understanding customer behavior. During global lockdowns, social media became a primary channel for expressing emotions, reshaping consumer behaviors, and disseminating public health information. Businesses utilized sentiment analysis to adapt strategies in real time, addressing customer concerns and responding to dynamic market conditions. By examining social media sentiment during the pandemic, studies revealed vital insights into public trust, economic recovery, and effective messaging strategies.

In the field of social media marketing analytics, tracking and analyzing customer sentiments and opinions of particular brands, goods, or services that are linked to the consumer generated content on social media is a current trend (Hemann & Burbry, 2013). Since consumer voices influence in building brand equity, analysis of consumers comments or opinions on Twitter can guide the top-level management to identify consumers' insight regarding product features, prices, promotion, sales channels and adopt right strategies to increase organizational performance. Sentiment analysis can help to identify how company's product is perceived by the users and to seize new opportunities regarding the improvement of products and services. By analyzing customer sentiment expressed in tweets with a product-related hashtag, marketers can quickly gather consumer feedback on a new product (Dhaoui, C., Webster, C. M., & Tan, L. P., 2017).

In the case of sentiment analysis, the Machine Learning is considered as the most accurate approach which has been used in marketing research as well (Pang, B. et al, 2002, Chaovalit & Zhou, 2005; Pathak, 2017). The term "sentiment" refers to the tracking of predictive judgments and the automatic analysis of evaluative text (Pang & Lee, 2008). It was initially coined by Das and Chen (2001) on the analysis of marketing sentiment. Though sentiment analysis has been extensively studied and used in many studies, there isn't much research that shows how marketers may use sentiment analysis to identify changing trends and changes in customer preferences for various product categories. Moreover, the majority of earlier studies that used sentiment analysis using Machine Learning did so by measuring the impact of customer engagement on product or service sales using quantitative data that was typically gathered through questionnaires (Ahire, V., Borse, S. 2022; Sailunaz & Alhajj, 2019; Chevalier and Mayzlin, 2006). Businesses can gain a deeper understanding of their customer base and their emotions through sentiment analysis. This understanding helps with marketing message optimization, trend forecasting, product and service development, and online reputation monitoring. Businesses can provide a product or service that best meets the demands of their customers by incorporating the customer's perspective into their offerings and business dealings (Grljević, O., & Bošnjak, Z. (2018).

Due to its distinct benefits, sentiment analysis is becoming more and more common in research. First, computer-aided sentiment analyses are not only more effective than manual coding, but they also yield results that are comparable. In their analysis of 800 travel evaluations from previous farm stay guests, Capriello et al. (2013) compare the effectiveness of two computer-aided sentiment analysis algorithms with manual content coding. They discovered that the outcomes of all three analyses are equally trustworthy. Second, by employing automatic algorithms to sift through text, sentiment analysis can efficiently minimize cost, time, and manual labor when compared to traditional approaches (such as surveys or focus groups) (Chiu et al., 2015).

In this paper, we explore the use of Machine Learning tool to analyze Tweet data to detect user trends from marketing perspective, and our objective is to detect evolving patterns and shifts in consumer preferences across product categories.

Since Twitter is considered as the most popular and common platforms in social media where people can express their opinions, thoughts, views, preferences freely, in our study we selected Twitter for our purpose. In the field of marketing sentiment analysis research is becoming more and more important with the increasing marketing efforts on more customization and ingenious offers targeting customers' need.

This study builds upon these advancements by applying a robust framework to analyze sentiment trends, consumer behaviors, and thematic patterns within social media data. Using Twitter as the focal platform and "car" as an illustrative product category, this research demonstrates the potential of machine learning-driven sentiment analysis to inform strategic decision-making across industries.

As discussed above, sentiment analysis can be used in marketing to find out whether customers like or dislike a product or to assess a company's reputation. Based on the data, a business can modify its marketing plan, make new product investments, improve the offering, etc. It can also be used to take instantaneous/fast actions based on their results.

Sentiment analysis and machine learning have come a long way, although this work can still address a lot of outstanding research issues and problems. The following important academic topics are still mostly unanswered and present worthwhile chances for investigation. In line with suggestions given by Mehraliyev, F., Chan, I. C. C., & Kirilenko, A. P. (2022), we formulate the following research questions for this study.

1) How can sentiment analysis help organizations identify user trends in a particular product category for taking strategic marketing decisions?
2) How can automated systems be designed to better identify and understand irony and sarcasm in social media content while successfully integrating live sentiment analysis with trend prediction from real-time data streams?

In order to answer these research questions, we set the following research objectives. The primary aim of this study is to develop a robust and scalable framework for analyzing consumer sentiment, extracting related product mentions "for example - car", and uncovering thematic patterns from large-scale social media data. This framework seeks to address gaps in existing sentiment analysis methodologies by leveraging advanced machine learning techniques to provide actionable insights for businesses.

The specific objectives of this research are:

a) To examine sentiment variations in product-related tweets, focusing on text lengths, word counts, and temporal dimensions (hours, days, and months) to identify emerging trends and behavioral patterns.
b) To identify related product terms and entities using advanced techniques such as Named Entity Recognition (NER) and semantic similarity analysis, enabling a deeper understanding of product associations and consumer concerns.

c) To employ topic modeling methods, such as Latent Dirichlet Allocation (LDA), to uncover hidden themes and discussion contexts in consumer tweets, providing insights into broader emotional and functional aspects of product discourse.

d) To propose a scalable and generalizable methodology that can be applied across diverse product categories and industries, with "car" serving as a representative example.

e) To address challenges such as detecting nuanced emotions (e.g., sarcasm and irony), handling the dynamic nature of real-time social media data, and ensuring adaptability to multilingual datasets.

The structure of this paper is organized to ensure clarity and coherence. The Literature Review provides a concise overview of related works relevant to the study. The Materials and Methodology section details the dataset utilized in this research and presents the proposed methodology employed for analysis. In the Results and Discussion section, we offer an in-depth presentation of the results obtained, followed by a critical discussion of their implications. The Limitations section addresses the challenges and constraints encountered during the research process. Finally, the Conclusions summarize the findings and highlight their significance, providing a comprehensive overview of the research.

## Literature review

Sentiment analysis has experienced a dramatic upheaval, especially since social media became a veritable gold mine of customer feedback. Support Vector Machines (SVM) and Naive Bayes classifiers were two of the early pioneers in this discipline who made a name for themselves by accurately classifying emotions. On sites like Twitter, however, these techniques frequently run into the difficulties of human language, such irony and sarcasm.

Salutations from the era of deep learning. The landscape transformed significantly with the advent of Recurrent Neural Networks (RNNs) and their more sophisticated counterpart, Long Short-Term Memory (LSTM) networks. Like remembering the story turns into an engrossing book, these models could now understand and retain the sequential structure of text. Still, a number of contentious problems have not been handled.

Analyzing multilingual data is one of the urgent problems. Bidirectional Encoder Representations from Transformers, or BERT, have improved context awareness by reading words backwards. But using these models in many languages is like trying to balance several hot torches at once difficult and prone to error. New approaches that aim to close these linguistic gaps include multilingual embeddings and transfer learning methods as XLM-R (Alexis Conneau, 2019) and mBERT (Jacob Devlin, 2018).

The cunning creatures of social media, sarcasm and irony frequently flip the meaning of a remark. It's a fascinating riddle even although some have attempted to control these with hybrid methods and context-aware models (Aniruddha Ghosh, 2016; Aditya Joshi, 2015). The requirement for real-time sentiment analysis adds to this complexity; picture attempting to build a ship as it is sailing. Online learning techniques and technologies like Apache Kafka and Flink (Paris Carbone, 2015) are advancing this field and allowing us to process and evaluate live data streams more quickly. More complication is introduced by the emergence of multimodal material, in which text, photos, and videos mix together like guests at a virtual cocktail party. Recent advances in multimodal neural networks, such Vision Transformer (ViT) (Alexey Dosovitskiy, 2020) and VideoBERT (Chen Sun, 2019), are advancing sentiment analysis by combining these many data sets. The trip is far from done, though.

Like the transient mutterings on Twitter, short texts have their own set of difficulties because of their concision and lack of context. To better comprehend and analyze these little bits of information, researchers are investigating advanced contextual embeddings like ELMo (Justyna Sarzynska-Wawe, 2021) and the robust capabilities of models like GPT-3 (Tom Brown, 2020).
In conclusion, sentiment analysis is still a difficult field even if we have come a long way. Every challenge from multilingual data to the nuances of sarcasm, real-time processing, multimodal material, and brief text analysis offers a chance for creativity and discovery. By filling in these gaps, we can improve our instruments and our grasp of customer mood, giving companies the information, they need to successfully negotiate the always-changing social media scene.

## Materials and Methodology

This study employs the **Sentiment140 dataset**, consisting of 1.6 million labeled tweets (positive and negative sentiments), which serve as the foundation for analyzing consumer discussions about various products. Tweets mentioning the keyword **"car"** were filtered, resulting in a subset of 41,387 tweets. The keyword **"car"** is used as an example product category to demonstrate the methodology. The analysis involves **data preprocessing**, **sentiment analysis**, **word association analysis**, and **topic modeling**. The methodology is organized as follows:

**5a) Dataset and Data Collection:**
- **Dataset:** The **Sentiment140 dataset** contains tweets labeled as positive (4) and negative (0).
- **Keyword Filtering**: Tweets mentioning the keyword "car" were extracted using case-insensitive filtering. This approach ensures relevance to the chosen product category while remaining adaptable to other keywords for future analyses.

**5b) Data Preprocessing:**
To prepare the data for analysis, several text-cleaning techniques were applied:
- **Noise Removal**: Removal of URLs, mentions, hashtags, special characters, and numerical values.
- **Lowercasing**: All text was converted to lowercase for uniformity.
- **Tokenization**: Texts were split into individual words using lightweight tokenization methods.
- **Stopword Removal**: Common stopwords were removed to retain meaningful words.

**5c) Sentiment Analysis:**
- **Text Length and Word Count Analysis**: Tweets were categorized into **Very Short**, **Short**, **Medium**, and **Long** based on text length and word count. Sentiment counts were analyzed across these categories.
- **Temporal Sentiment Trends**: Sentiment distributions were analyzed over **hours of the day**, **days of the week**, and **months of the year** to identify time-based variations.

**5d) Word Association Analysis**

- **Word Frequency Analysis**: The **TF-IDF (Term Frequency-Inverse Document Frequency)** technique was used to extract the top 20 most frequent product-related words (e.g., *car*, *phone*, *laptop*, etc.).
- **Semantic Similarity**: Pre-trained **Word2Vec** and **Sentence Transformers** models were employed to identify the top 10 words semantically similar to the keyword "car".
- **Named Entity Recognition (NER)**: A fine-tuned **BERT-based NER model** identified product-related entities such as brands (*Tesla*, *BMW*) and categories (*vehicle*, *sedan*).

**5e) Topic Modeling**

- **LDA (Latent Dirichlet Allocation)**: Topic modeling was performed to identify latent themes within the filtered tweets. The process involved:
    - **Text Representation**: Conversion of preprocessed tweets into **bag-of-words** format using a dictionary.
    - **Modeling**: An **LDA model** with 4 topics was trained to uncover hidden themes in discussions about *cars*.
    - **Topic Visualization**: The top keywords within each topic were visualized to interpret discussion themes.

**5f) Tools and Libraries**

The implementation of the study was done using Python with the following libraries:
- **Data Handling**: pandas, numpy
- **Preprocessing**: re, nltk, sklearn
- **Word Analysis**: TfidfVectorizer, gensim for word embeddings and LDA
- **NER**: Hugging Face **Transformers** (dbmdz/bert-large-cased-finetuned-conll03-english)
- **Visualization**: matplotlib, seaborn

**5g) Output and Validation**

The analysis results were validated using:
- **Sentiment Trends**: Examining sentiment variations across text length and temporal dimensions.

- **Semantic Relationships**: Identifying contextual relationships among keywords using word embeddings.
- **Topic Interpretability**: Ensuring the coherence and quality of topics generated by LDA for actionable insights.

This structured methodology ensures scalability and adaptability for analyzing sentiment across diverse product categories and industries.

## Results and Discussion

The results provide a comprehensive analysis of consumer sentiment, word associations, and thematic patterns regarding product-related discussions on Twitter, using "car" as an example product.

**a) Sentiment Analysis Results**

The sentiment distribution for tweets mentioning the keyword "car" revealed that **negative sentiments (58.2%)** significantly outnumbered **positive sentiments (41.8%)**.

- **Text Length Categories**: Sentiment counts were highest for *short and medium* tweets, with negative sentiment dominating across all text-length categories.
- **Temporal Analysis**:
    - **Hour of the Day**: As we can see from fig. (a) negative sentiment peaked around **7 AM** and **7 PM**, while positive sentiment remained relatively stable but lower throughout the day.
    - **Days of the Week**: Negative sentiment surged on **Saturday (1,232 tweets)** and **Sunday (981 tweets)**, indicating increased user activity and discussions on weekends, which is indicating in fig. (b).
    - **Months of the Year**: By the fig. (c) we can see that a notable rise in discussions was observed from **April to June**, with negative sentiment increasing significantly, suggesting seasonal trends in product mentions.

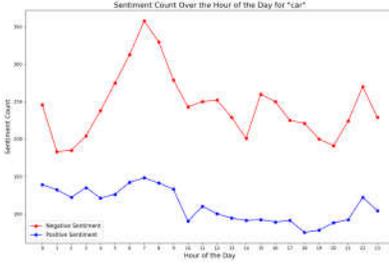 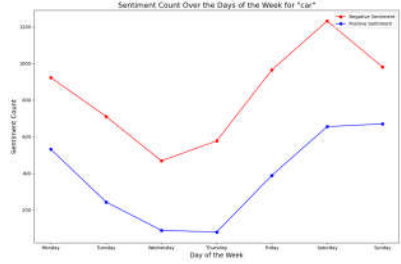 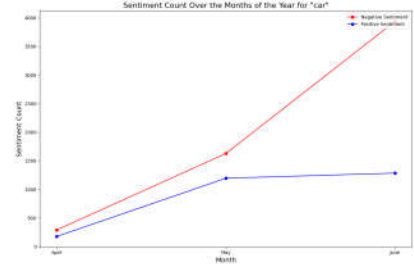

|  (a)  |  (b)  |  (c)  |

### b) Model Performance:

Machine learning models were evaluated based on their ability to classify sentiment. Among the models tested, BERT demonstrated the best performance, achieving:

- Accuracy: 83.48%
- Precision: 79.37%
- Recall: 90.60%
- F1 Score: 84.61%

| Model | Precision (%) | Recall (%) | Accuracy (%) | F1 Score (%) |
|---|---|---|---|---|
| SVM | 70.85 | 76.25 | 75.43 | 73.45 |
| Naive Bayes | 76.38 | 75.74 | 76.10 | 76.06 |
| LSTM | 79.42 | 84.14 | 81.12 | 81.71 |
| **BERT** | **79.57** | **90.60** | **83.48** | **84.61** |

The superior performance of BERT highlights its ability to handle complex linguistic patterns and contextual relationships, making it particularly effective for sentiment classification.

**c) Word Association Analysis:**

- **Top 20 Product-Related Words:** The most frequent words included watch, phone, car, iphone, and laptop, highlighting strong associations with consumer products.

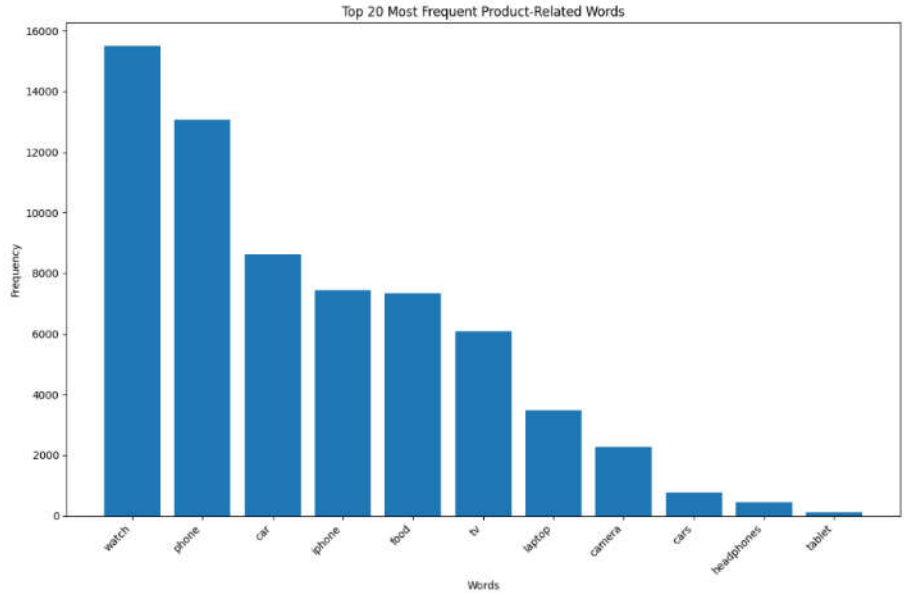

- **Semantic Similarity:** Using Word2Vec embeddings, words like drive, accident, insurance, and keys showed high semantic similarity to "car", reflecting user concerns around driving, safety, and ownership experiences.

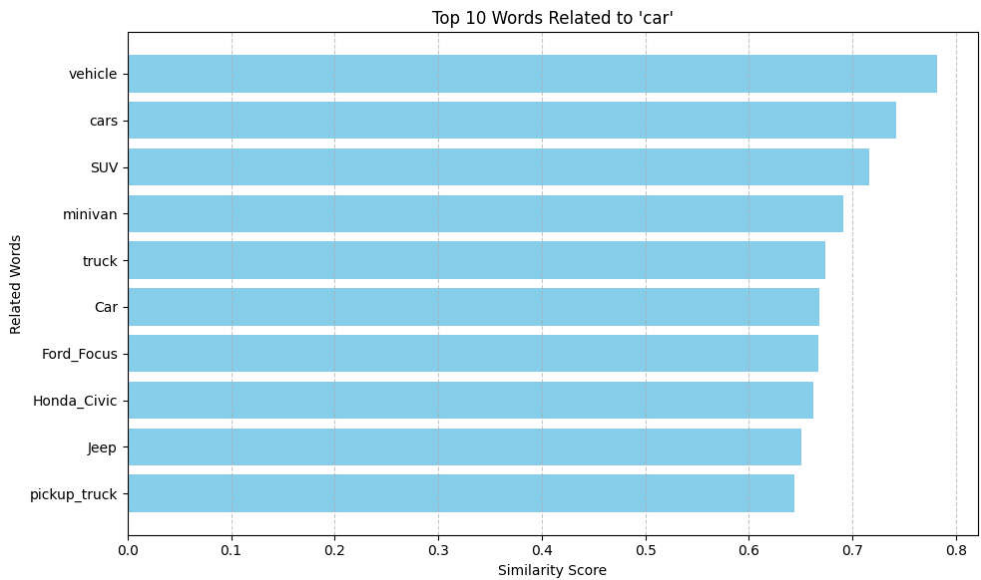

- **Named Entity Recognition (NER):** The NER pipeline identified brands such as Tesla, BMW, and generic terms like vehicle and sedan, demonstrating the dataset's capability to capture product-specific mentions.

### d) Topic Modeling Results

The **LDA-based topic modeling** uncovered four key themes within tweets about "car":

- **Topic 0**: Emotion and entertainment discussions, including keywords like *scary*, *movie*, *watching*, and *funny*.
- **Topic 1**: Personal experiences and ownership, with frequent words like *car*, *got*, *her*, and *just*.
- **Topic 2**: Maintenance and care-related discussions, featuring terms like *take*, *care*, and *new*.
- **Topic 3**: Broader sentiments and opinions, including *you*, *the*, *and*, and *have*, indicating generic user expressions.

| Topic | Word 1 | Word 2 | Word 3 | Word 4 | Word 5 | Word 6 | Word 7 | Word 8 | Word 9 | Word 10 |
|---|---|---|---|---|---|---|---|---|---|---|
| Topic 0 | scary (0.013) | card (0.013) | watching (0.011) | movie (0.009) | follow (0.007) | funny (0.007) | scared (0.006) | sounds (0.006) | cartoon (0.006) | watch (0.005) |
| Topic 1 | the (0.056) | and (0.039) | car (0.032) | with (0.018) | her (0.013) | just (0.013) | got (0.011) | was (0.010) | she (0.008) | for (0.008) |
| Topic 2 | car (0.036) | the (0.031) | take (0.023) | for (0.022) | care (0.019) | and (0.017) | get (0.013) | day (0.010) | have (0.010) | new (0.010) |
| Topic 3 | you (0.038) | the (0.035) | and (0.019) | that (0.018) | for (0.014) | but (0.013) | have (0.012) | care (0.012) | your (0.011) | dont (0.010) |

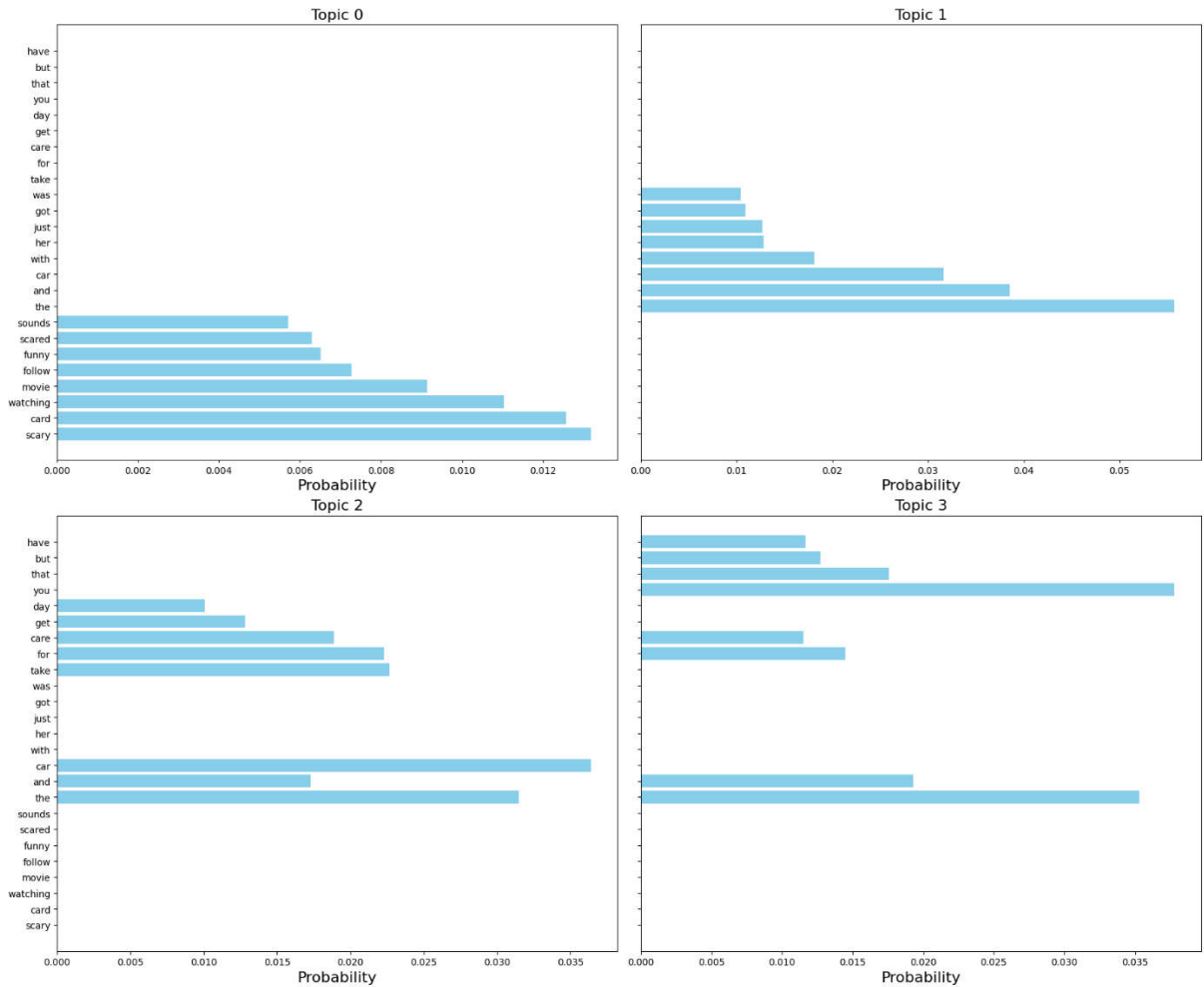

The results revealed that negative sentiments, particularly in short and medium-length tweets, often reflect consumer dissatisfaction with issues like repairs, accidents, or maintenance. Temporal trends identified peak user activity on weekends and during specific hours, enabling organizations to:

- **Optimize Engagement**: Target users during high-activity periods with relevant campaigns.
- **Address Concerns**: Focus on common pain points like safety and insurance to build customer trust.

The analysis of tweets mentioning the keyword "car" revealed key insights into consumer sentiment, trends, and behavioral patterns. **Sentiment analysis results** show that negative sentiments (58.2%) dominate discussions, particularly in *short and medium* tweets, indicating user frustration or dissatisfaction regarding issues such as repairs, accidents, or maintenance. Temporal analysis highlights specific periods of heightened activity, such as **weekends** (highest on Saturday with 1,232 negative tweets) and **mornings/evenings** (peak hours of 7 AM and 7 PM). This enables organizations to:

- Identify when users are most active and likely to express concerns or positive experiences.
- Develop targeted marketing strategies to address negative perceptions, such as weekend promotions or content tailored to common issues like repairs and insurance.

Additionally, **word association results** (e.g., *drive*, *accident*, *insurance*, *keys*) and **NER-based entity recognition** (e.g., *Tesla*, *BMW*, *vehicle*) highlight specific aspects of product usage and associated brands. These insights help businesses understand which features or issues are most discussed and strategically address user needs in marketing campaigns.

|  | precision | recall | f1-score | support |
|---|---|---|---|---|
| negative | 0.89 | 0.76 | 0.82 | 4539 |
| positive | 0.79 | 0.91 | 0.85 | 4564 |
| accuracy |  |  | 0.83 | 9103 |
| macro avg | 0.84 | 0.83 | 0.83 | 9103 |
| weighted avg | 0.84 | 0.83 | 0.83 | 9103 |

The Classification Report provided a detailed breakdown of performance across positive and negative sentiments, with high recall indicating the model's strength in correctly identifying the majority of negative sentiments, which dominate the dataset.

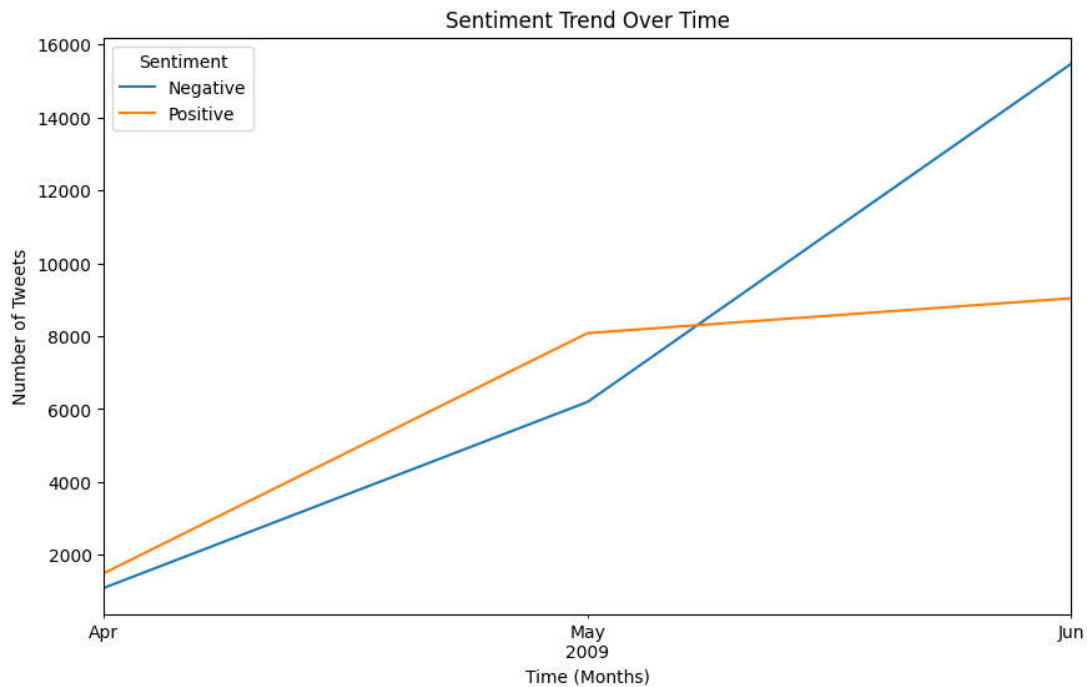

The temporal analysis further revealed that negative sentiments are highest on weekends and peak during morning and evening hours. Word clouds for positive and negative tweets highlighted terms such as "drive", "accident", and "insurance" in negative contexts and "new", "good", and "love" in positive contexts. Organizations can leverage these insights to identify user pain points, such as accidents or insurance concerns, and align marketing efforts or customer engagement strategies to address these issues during peak activity periods.

The inclusion of **Topic Modeling (LDA)** helped identify thematic patterns that often contain subtle sentiments or sarcasm. For instance:

- **Topic 0**: Emotion-driven discussions (*scary*, *funny*, *watching*) suggest content that might involve humor, sarcasm, or entertainment contexts.

- **Topic 2**: Keywords like *care*, *take*, and *new* indicate possible sarcasm around car maintenance. For example, phrases such as "*taking care of my car again, how fun!*" can be sarcastic despite seemingly positive words.

These findings emphasize the need for **context-aware sentiment analysis tools** that can detect linguistic nuances like irony and sarcasm using advanced natural language processing techniques. This can improve the accuracy of live sentiment pipelines, ensuring organizations gain a deeper understanding of consumer attitudes in real time.

Furthermore, **semantic similarity analysis** (using Word2Vec) captured words closely associated with "car" (e.g., *drive*, *insurance*, *keys*, *fixed*), showcasing the model's ability to identify related terms automatically. This enhances trend prediction pipelines by dynamically incorporating emerging topics, words, and user behaviors.

The model performance metrics demonstrate a robust ability to detect sentiment polarity; however, challenges remain in detecting sarcasm and irony. Topic Modeling (LDA) and **semantic similarity analysis** addressed this by capturing hidden themes and related terms:

- **Topic 0** included words like *"scary"*, *"funny"*, and *"watching"*, which suggest humor or sarcasm-driven discussions.
- **Top Similar Words** (using Word2Vec) for "car" included terms like *"drive"*, *"insurance"*, *"wash"*, and *"hit"*, offering context to identify sarcasm.

These insights demonstrate that integrating advanced **context-aware models** (e.g., Transformer-based systems) can enhance automated pipelines by detecting subtle sentiment cues. Further integration with **real-time trend prediction** can dynamically update models to reflect emerging themes and user behaviors.

In conclusion, sentiment analysis and advanced methods like semantic similarity, NER, and topic modeling collectively answer **RQ1 and RQ2**, demonstrating the potential to extract actionable trends, identify subtle sentiment cues, and improve real-time monitoring of consumer perceptions. This approach is highly adaptable to other product categories beyond "car", providing a robust framework for organizations to inform strategic marketing decisions and enhance automated systems for trend prediction.

This study extends the limited research on sentiment analysis for predicting consumer behavior in a certain product category by introducing a new methodology. This study reveals a novel methodology for collecting comprehensive data and preparing workflow to cleanse and structure Twitter data for analysis. Advanced machine learning models including Support Vector Machines (SVM), Naive Bayes, Long Short-Term Memory (LSTM) networks, and Bidirectional Encoder Representations from Transformers (BERT) were utilized to classify sentiments and predict trends. This methodology enhances prediction accuracy in one hand, on the other hand it provides deeper insights into consumer behavior.

By tackling issues including sarcasm identification, multilingual data processing, and the incorporation of real-time analysis, this study makes a substantial contribution to sentiment analysis research. This research makes it easier to get accurate and useful consumer insights by offering a scalable framework that can be applied to a variety of industries. This helps firms to successfully navigate the ever-changing social media-driven market environment.

## Limitations

While this study presents a robust framework for analyzing consumer sentiment and product discussions, it is not without limitations. Firstly, the analysis relies on the Sentiment140 dataset, which is constrained to three months of Twitter data, potentially excluding seasonal or long-term trends in consumer sentiment. Additionally, the dataset focuses solely on English-language tweets, limiting the generalizability of findings to multilingual contexts or regions with diverse linguistic characteristics.

Detecting nuanced emotions such as sarcasm, irony, and implicit sentiment remains a significant challenge, as these require deeper contextual understanding that may not always be captured by existing models. The domain-specific focus on the keyword "car" further narrows the scope of this research, necessitating validation across other product categories to confirm the scalability and adaptability of the proposed methodology.

Moreover, the static nature of the dataset prevents real-time sentiment analysis, which is crucial for applications requiring immediate insights. Future studies could address this by integrating live data streams and developing online learning models. Finally, while advanced models like BERT demonstrated superior performance, their computational demands can pose challenges for smaller organizations or real-time deployments, where computational resources may be constrained.

Despite these limitations, this research provides a solid foundation for exploring consumer sentiment trends, with potential to expand its applicability through real-time pipelines, multimodal datasets, and multilingual analysis frameworks.

## Conclusions

This study presents a comprehensive framework for analyzing consumer sentiment and product discussions on social media using advanced machine learning and NLP techniques. By leveraging the Sentiment140 dataset, the research effectively demonstrates how sentiment analysis, word association, and topic modeling can uncover valuable insights into consumer preferences and behaviors. The superior performance of BERT underscores the potential of transformer-based models to handle complex and contextual sentiment classification tasks.

Key findings include the dominance of negative sentiment across textual and temporal dimensions, identification of semantically related terms such as "drive" and "insurance," and thematic patterns that reveal broader emotional and functional contexts of product discussions. These insights are invaluable for businesses aiming to optimize marketing strategies, address consumer concerns, and improve customer engagement through data-driven decision-making.

Despite its notable contributions, the study acknowledges several limitations. The reliance on the Sentiment140 dataset, constrained to three months of English-language Twitter data, may exclude long-term trends and multilingual perspectives. Challenges in detecting nuanced sentiments such as sarcasm and irony further underscore the need for more context-aware models. Additionally, the computational demands of advanced models like BERT pose accessibility challenges for smaller organizations or real-time applications. Addressing these limitations, future research will focus on integrating live data streams, incorporating multimodal and multilingual datasets, and exploring adaptive models for real-time sentiment analysis.

Overall, this research provides a scalable and adaptable approach to sentiment analysis, offering a robust foundation for understanding consumer behavior in an increasingly digital and dynamic marketplace. The proposed methodology can be extended across diverse product categories and industries, equipping organizations with the tools to derive actionable insights and remain competitive in the fast-evolving landscape of social media analytics.